\documentstyle[12pt]{article}

\def\Journal#1#2#3#4{{#1} {\bf #2}, #3 (#4)}

\def\NPB{{\em Nucl. Phys.} B}

\def\PRD{{\em Phys. Rev.} D}

\begin{document}

\begin{flushright}
NEIP-98-016\\
hep-th/9811127 \\
November 1998
\end{flushright}

\bigskip
\begin{centering}
\bigskip
{\bf ON THE RELATION BETWEEN WILSON ACTION\\ 
AND CJT EFFECTIVE ACTION\footnote{Talk given by G.A.-C.
at the {\it 6-th International Conference on Path Integrals 
from peV to TeV}.
To appear in the 
proceedings.}}

\bigskip
\bigskip
\medskip

{\bf Giovanni AMELINO-CAMELIA} and {\bf Jean-Pierre DERENDINGER}

\bigskip

Institut de Physique, Universit\'e de Neuch\^atel,
2000 Neuch\^atel, Switzerland

\end{centering}
\vspace{1cm}

\begin{abstract}
A relation between Wilson action and
the Cornwall-Jackiw-Tomboulis effective action, which was recently
suggested 
by Periwal, is here derived using path-integral techniques.
\end{abstract}

\bigskip

The Wilson action\cite{wilson} and the CJT (Cornwall-Jackiw-Tomboulis)
effective action\cite{cjt} have several applications in 
theoretical physics,
especially for the analysis of certain non-perturbative aspects
of quantum field theories.
A noticeable example of the usefulness of the
Wilson action is provided by its role in the recent
understanding of certain non-perturbative aspects
of supersymmetric quantum field theories,
and in those studies the holomorphicity properties
of the Wilson action have been very important \cite{holo}.
Similarly, the usefulness of the CJT effective
action for studies of dynamical mass generation \cite{cjt,dynamass}
and temperature induced phase transitions \cite{gacpi}
is due mostly to some specific properties of this action,
notably the fact that under certain conditions \cite{bara,banf}
it can be interpreted as the free energy in an ensemble with prescribed 
field expectation value  and prescribed connected two-point correlation 
function and the fact that it admits a loop expansion, even though
it captures some non-perturbative (resummed perturbative)
aspects of the theory.
Motivated by the hope to eventually be able to combine in a single structure
the formidable properties of the Wilson action and the CJT effective
action, in a recent paper \cite{peri} Periwal 
has suggested a relation between these two actions.
We shall first review, to the extent necessary for our analysis, 
the definitions of the Wilson action and the CJT effective
action, then we shall remind the reader of the relation
suggested by Periwal,
and finally we shall derive this relation using 
path-integral techniques.

For simplicity, we limit our discussion
to the case of a theory with two scalar fields $\Phi$ and $\Sigma$
and Lagrangian $L(\Phi,\Sigma)$. For most of the analysis the reader
is invited to think of $\Phi$ and $\Sigma$ respectively as the 
light and heavy modes of a single field (where light modes
are those below a given cut-off scale $\Lambda$ while the heavy modes
are above $\Lambda$).
In such a context, 
the Wilson action $S^W$ can be formally introduced as \cite{peri}
\begin{equation}
S^W(\Phi) = - i \log \int {\cal D} \Sigma \, \exp \left[ i \int dx \,
L(\Phi,\Sigma) \right] ~.
\label{swdef}
\end{equation}
Besides this ordinary viewpoint in which $\Phi$ and $\Sigma$
represent light and heavy modes of a single field,
this formal definition is sometimes useful \cite{peri}
also in contexts in which $\Phi$ and $\Sigma$
represent two unrelated fields. Therefore, in the following
we shall not explicitly refer to the $\Lambda$-dependence
introduced by the separation between $\Phi$-modes and $\Sigma$-modes. 

In order to define the CJT effective action $\Gamma^{CJT}$
one first introduces local and bilocal sources
\begin{equation}
\begin{array}{rcl}
Z[J_\Phi,J_\Sigma,K_\Phi,K_\Sigma] 
\!\! &=& \!\! \int {\cal D} \Phi \, {\cal D} \Sigma  
\, \exp [ i \int_x L(\Phi,\Sigma)
+i \int_x (J_\Phi \Phi + J_\Sigma \Sigma) \\
&&{}~~~~~~~~~~~~~~~~~
+i \int_{x,y} (\Phi K_\Phi \Phi/2 + \Sigma K_\Sigma \Sigma/2) ]~,
\end{array}
\label{wcjtdef}
\end{equation}
where we adopted a compact notation
such that, {\it e.g.}, $\int_{x,y} \equiv \int dx \, dy$, 
$J_\Phi \Phi \equiv J_\Phi(x) \Phi(x)$ and
$\Phi K_\Phi \Phi \equiv \Phi(x) K_\Phi(x,y) \Phi(y)$.

{}From $Z[J_\Phi,J_\Sigma,K_\Phi,K_\Sigma]$ the CJT
effective action $\Gamma^{CJT}[\phi,\sigma,G_\phi,G_\sigma]$
is obtained via Legendre transform:
\begin{equation}
\begin{array}{rcl}
\Gamma^{CJT}[\phi,\sigma,G_\phi,G_\sigma] \!\!\! &=& \!\!\!
-i \log Z[J_\Phi,J_\Sigma,K_\Phi,K_\Sigma] -
\int_x (J_\Phi \phi + J_\Sigma \sigma) \\
&&{}\!\!\!\!\!\!
- \int_{x,y} (\phi K_\Phi \phi/2 + \sigma K_\Sigma \sigma/2
+G_\phi K_\Phi/2 + G_\sigma K_\sigma/2) ~,
\end{array}
\label{cjtdef}
\end{equation}
where $\phi$,$\sigma$,$G_\phi$,$G_\sigma$ are such that
\begin{equation}
{\delta \Gamma^{CJT} \over \delta \phi} = 
- J_\Phi - \int_y K_\Phi \phi ~,
\label{defa} 
\end{equation}
\begin{equation}
{\delta \Gamma^{CJT} \over \delta \sigma} = 
- J_\Sigma - \int_y K_\Sigma \sigma ~,
\label{defb} 
\end{equation}
\begin{equation}
{\delta \Gamma^{CJT} \over \delta G_\phi} = 
- K_\Phi /2 ~,
\label{defc} 
\end{equation}
\begin{equation}
{\delta \Gamma^{CJT} \over \delta G_\sigma} = 
- K_\Sigma /2 ~.
\label{defd} 
\end{equation}
As mentioned, 
for practical applications it is important that $\Gamma^{CJT}$
admits the expansion  \cite{cjt}:
\begin{equation}
\begin{array}{rcl}
\Gamma^{CJT} &=& V_{tree}(\phi,\sigma)
+ {i \over 2} \, \int_k \, [D_{\phi}^{-1}(\phi;k) G_\phi(k)
+ D_{\sigma}^{-1}(\phi;k) G_\sigma(k)]\\
&&{} + {i \over 2} \, \int_k \, 
[\ln G_\phi^{-1}(k) +\ln G_\sigma^{-1}(k) ] 
+ \Gamma_{2L}[\phi,\sigma;G_\phi(k),G_\sigma(k)]
~,\label{hfb}
\end{array}
\label{cjtloop}
\end{equation}
where $V_{tree}$ is the tree-level (classical) potential, 
$D_{\phi},D_{\sigma}$ are the tree-level propagators \cite{cjt},
and $\Gamma_{2L}$ is given by 
all two-particle-irreducible
vacuum-to-vacuum
graphs with two or more loops
in the theory with vertices given by the interaction part of the 
shifted ($\Phi \rightarrow \Phi + \phi$,
$\Sigma \rightarrow \Sigma + \sigma$) Lagrangian and
propagators set equal to $G_\phi$,$G_\sigma$.

We are here concerned with Periwal's suggestion \cite{peri}
of the following additional property of $\Gamma^{CJT}$:
\begin{equation}
S^W(\phi) = \Gamma^{CJT} [\phi, \sigma = \sigma^o, 
G_\phi=0, G_\sigma=G_\sigma^o] ~,
\label{periconj} 
\end{equation}
where $\sigma^o$ and $G_\sigma^o$
are such that $J_\Sigma = 0 = K_\Sigma$
for $\sigma \! = \! \sigma^o$ and
$G_\sigma \! = \! G_\sigma^o$ (see Eqs.~(\ref{defb}) and (\ref{defd})).

In order to verify the relation (\ref{periconj})
we start by observing that from Eqs.~(\ref{wcjtdef}) and (\ref{cjtdef})
one can derive
\begin{equation}
\begin{array}{rcl}
\Gamma^{CJT} [\phi, \! \sigma, \! G_\phi, \! G_\sigma]
\!\!\!&=& \!\!\! - i \log \int \! {\cal D} \Phi \, {\cal D} \Sigma  
\, \exp \{ i \int_x L(\Phi,\Sigma) +i \int_x J_\Phi (\Phi \! - \! \phi) \\
&&{}
\!\!\!
+(i/2) \int_{x,y} [\Phi K_\Phi \Phi - \phi K_\Phi \phi 
- K_\Phi G_\phi ] +i \int_x J_\Sigma (\Sigma \! - \! \sigma) \\
&&{}
\!\!\!
+(i/2) \int_{x,y} [\Sigma K_\Sigma \Sigma - \sigma K_\Sigma \sigma
- K_\Sigma G_\sigma ] \}~.
\end{array}
\label{stepone}
\end{equation}
Next we use Eqs.~(\ref{swdef}) and (\ref{defa})
and the fact that $J_\Sigma = 0 = K_\Sigma$
for $\sigma \! = \! \sigma^o$ and
$G_\sigma \! = \! G_\sigma^o$ to obtain
\begin{equation}
\begin{array}{rcl}
\Gamma^{CJT} [\phi, \! \sigma \! = \! \sigma^o, \! 
G_\phi, \! G_\sigma \! = \! G_\sigma^o]
\!\!\!  &=& \!\!\! - i \log \int \! {\cal D} \Phi 
\, \exp \{ i S^W(\Phi)
-i \int_x (\delta \Gamma^{CJT} / \delta \phi) (\Phi \! - \! \phi) 
\\
&&{}\!\!\!\!\!
+ (i/2) \int_{x,y} [K_\Phi (\Phi - \phi)^2 - K_\Phi G_\phi] \}~.
\end{array}
\label{steptwo}
\end{equation}
The only element of Periwal's proposal
which we have not yet implemented is $G_\phi =0$.
Within our path-integral approach 
one should take into account
that $G_\phi \rightarrow 0$ corresponds 
to $K_\Phi \rightarrow i \infty$. This property, which
can be verified directly by using Eqs.~(\ref{defc}) 
and (\ref{cjtloop}),
is related to the fact \cite{peri} 
that $G_\phi$ plays the role of propagator for the 
expansion of $\Gamma^{CJT}$ while $K_\Phi$, 
as seen in Eq.~(\ref{wcjtdef}), is introduced as 
a mass for the field $\Phi$.
From Eq.~(\ref{steptwo}) one sees that as $K_\Phi \rightarrow i \infty$
the integrand of the $\Phi$-functional integration
becomes more and more
sharply peeked around configurations $\Phi \sim \phi$.
In the limit $K_\Phi = i \infty$ the
term $(\delta \Gamma^{CJT} / \delta \phi) (\Phi \! - \! \phi)$
can be dropped and the only dependence of the path integral
on $S^W$ is through $S^W(\phi)$, which is also the
only $\phi$-dependent element of the path integral.
Therefore, up to an irrelevant additive $\phi$-independent
contribution, we obtain the relation (\ref{periconj})
from (\ref{steptwo}).

Our analysis, within the limitations of a formal path-integral derivation,
provides support for Periwal's suggestion.
As already emphasized by Periwal, some of the possible applications
of this relation might rely on non-perturbative convexity
properties of the CJT effective action, which would be transferred,
via  (\ref{periconj}), to the Wilson action.
It remains to be seen whether such convexity will be available
in the contexts of physical interest; in fact, it is well known
that in some cases \cite{gacpi,bara,banf}
singularities in the relevant Legendre transforms
spoil the convexity of the CJT effective action.

We also observe that,
since the conditions $\sigma \! = \! \sigma^o$, 
$G_\sigma \! = \! G_\sigma^o$ set in (\ref{periconj})
correspond to vanishing 
sources $J_\Sigma$, $K_\Sigma$,
one could express the Wilson action directly in terms 
of a CJT effective action obtained without 
ever introducing $J_\Sigma$ and $K_\Sigma$.
Denoting by ${\tilde \Gamma}^{CJT} [\phi, \! G_\phi]$
the CJT effective action obtained
via Legendre transform of
$Z[J_\Phi,J_\Sigma \! = \! 0,K_\Phi,K_\Sigma \! = \! 0]$,
one easily derives the relation $S^W(\phi) \!
= \! {\tilde \Gamma}^{CJT} [\phi, G_\phi \! = \!0]$.

\section*{Acknowledgments}
One of us (G.A.-C.) benefited from conversations
with Roman Jackiw.
This work was supported by
the Swiss National Science
Foundation.


\begin{thebibliography}{99}
\bibitem{wilson} K.G.~Wilson and J.B.~Kogut, 
{\em Phys. Rep.} C {\bf 12}, 75 (1974).

\bibitem{cjt}
J.M.~Cornwall, R.~Jackiw, and E.~Tomboulis,
\Journal{\PRD}{10}{2428}{1974}.

\bibitem{holo} N.~Seiberg and E.~Witten,
\Journal{\NPB}{426}{19}{1994};
\Journal{\NPB}{431}{484}{1994}.

\bibitem{dynamass} J.M.~Cornwall,
\Journal{\PRD}{26}{1453}{1982}.

\bibitem{gacpi} 
G.~Amelino-Camelia and S.-Y.~Pi,
\Journal{\PRD}{47}{2356}{1993}.

\bibitem{bara} 
T.~Banks and S.~Raby,
\Journal{\PRD}{14}{2182}{1976}.

\bibitem{banf}
R.~Jackiw and G.~Amelino-Camelia, hep-ph/9311324,
in {\it Proceedings of the Third Workshop on Thermal Field Theories
and Their Applications},
ed. F.C. Khanna, R.~Kobes, G.~Kunstatter, and
H.~Umezawa (World Scientific, Singapore, 1994).

\bibitem{peri}
V.~Periwal, \Journal{\PRD}{57}{6551}{1998}.

\end{thebibliography}
\end{document}